# Relationship between event counting statistics and waiting time statistics in the steady state

Seong Jun Park[a,b] and M.Y. Choi[b]

[a] National CRI-Center for Chemical Dynamics in Living Cells, Chung-Ang University, Seoul 06974, Korea

[b] Department of Physics and Astronomy and Center for Theoretical Physics, Seoul National University, Seoul 08826, Korea

**Abstract**

There are two main quantities involved in the deviation of a stochastic process from a Poisson process: the squared coefficient of variation of the time intervals between adjacent events and the Fano factor of the number of reaction events. As well known, these two quantities are equal for renewal processes, while their relationship remains unexplored for non-renewal processes. In this paper, we establish an explicit relation between the two statistics that is applicable to non-renewal processes. The new relation, which reduces to the previously mentioned result for renewal processes, is confirmed to be accurate in several cases of non-renewal processes.

# 1. Introduction

There are a variety of stochastic processes in natural and social phenomena that can be described in terms of a series of events. Examples include neuronal firing [1, 2], gene expression [3-5], chemical reactions [6, 7], earthquakes [8, 9], and human activities [10, 11]. The time intervals between events in the series are likely to be influenced by the environment in which the events occur, while the environment undergoes changes over time. In consequence, the probability density function of the inter-event times varies with each event occurrence; such a trend implies that the events do not represent a renewal process characterized by identical and independent inter-event time distributions. It is thus recognized that almost all event sequences in the universe are non-renewal in nature, making it crucial to understand non-renewal processes.

The behavior of a non-renewal process after a long period of time is of immense research interest, particularly in whether its behavior settles down probabilistically and in the relationship between the event counting statistics and the inter-event times. According to the renewal theory, the Fano factor of the number of events and the squared coefficient of variation of the inter-event times are equal in the steady state [12, 13]. Still, little attention has been paid to the relationship between the event counting statistics and the probability density function of the inter-event times for a non-renewal process.

In this work, we derive an equation satisfied by non-renewal processes. The relationship between the event counting statistics and the inter-event time distribution becomes manifest. The inter-event time distributions of non-renewal processes can vary with the occurrence of each event, while renewal processes have identical and independent inter-event time distributions. Considering the fluctuations of the inter-event time distribution in a non-renewal process, we investigate the probability density function of the time required to complete a given number of events and derive the equation satisfied by the waiting-time distribution and the event-counting statistics in the steady state.

This paper is organized as follows: In Section 2, we derive the relation between event-counting statistics and event-time statistics. In Section 3, this relationship is confirmed by the help of several exemplary non-renewal processes. Finally, Section 4 presents the summary and discussion.

## 2. General relation between event-counting statistics and event-time statistics in the steady state

In this section, we derive the relationship between event counting statistics and event time statistics in the steady state. The first step is to consider that the time taken for an event to occur $n$ times is denoted by $t_n = \sum_{i=1}^{n} x_i$. Its mean and second moment can be expressed as

$$\langle t_n \rangle = n\mu \tag{1-1}$$

$$\langle t_n^2 \rangle = n\alpha + n^2 \mu^2 \tag{1-2}$$

where $\mu \equiv \frac{1}{n}\sum_{i=1}^{n} \langle x_i \rangle$ and $\alpha \equiv \frac{1}{n^2}\sum_{i=1}^{n}\sum_{j=1}^{n} \langle x_i x_j \rangle - \langle x_i \rangle \langle x_j \rangle$ correspond to the sample mean and variance in statistics. To derive the relation between the event counting statistics and the waiting time distribution, we should pay attention to the probability density function of the waiting time until the event occurs $n$ times. To obtain the counting statistics of products in the steady state requires careful observation of its asymptotic behavior over a long period of time. The Laplace transform of a function is useful for its asymptotic expansion. Let $\varphi_n(t)$ denote the probability density function of the time required to produce at least $n$ products, and $\hat{\varphi}_n(s)$ denote the Laplace transform of $\varphi_n(t)$, i.e., $\hat{\varphi}_n(s) = \int_0^{\infty} \varphi_n(t) e^{-st} dt$. The power series expansion of $\hat{\varphi}_n(s)$ follows from $\langle t_n^m \rangle = (-1)^m \left. \frac{\partial^m \hat{\varphi}_n(s)}{\partial s^m} \right|_{s=0}$ and the normalization condition $\hat{\varphi}_n(0) = 1$:

$$\hat{\varphi}_n(s) = 1 + \sum_{m=1}^{\infty} \frac{(-1)^m \langle t_n^m \rangle}{m!} s^m \tag{2}$$

The long-time behavior of $\varphi_n(t)$ is governed by the behavior of $\hat{\varphi}_n(s)$ for small values of the Laplace variable $s$. Therefore, Eq. (2) is approximated by $1 - \langle t_n \rangle s + \frac{\langle t_n^2 \rangle}{2} s^2$. Substituting Eqs. (1-1) and (1-2) into this approximation and using $1 + x \approx e^x$ for small $x$ $(|x| \ll 1)$, we obtain

$$\hat{\varphi}_n(s) \approx e^{-n\langle t_1 \rangle s + \frac{n\alpha}{2}s^2} \qquad (3)$$

for small $s$.

We are now ready to obtain the counting statistics of products in a steady state. The relation between the counting statistics of products and the probability density function of the time to produce $n$ products is well established [13, 14].

$$\langle n(t) \rangle = \sum_{n=1}^{\infty} \int_0^t \varphi_n(\tau) d\tau \qquad (4\text{-}1)$$

$$\langle n(n-1)(t) \rangle = 2\sum_{n=1}^{\infty} (n-1) \int_0^t \varphi_n(\tau) d\tau, \qquad (4\text{-}2)$$

where $\langle n(t) \rangle$ is the mean and $\langle n(n-1)(t) \rangle$ is the second factorial moment of the product number at time $t$. Equations (3) and (4-1) allow us to obtain the Laplace transform of the mean product number:

$$\langle \hat{n}(s) \rangle = \frac{1}{s\left(e^{\langle t_1 \rangle s - \frac{1}{2}\alpha s^2} - 1\right)}. \qquad (5)$$

Expanding Eq. (5) in a power series for small values of $s$ and taking the inverse Laplace transform, we obtain the behavior of the mean product number for a long-time duration:

$$\langle n(t) \rangle \approx \frac{1}{\mu}t + \frac{1}{2}\left(\frac{\alpha}{\mu^2} - 1\right) \qquad (6)$$

for large $t$. In a similar manner, Eq. (4-2) together with Eq. (3) leads to the second factorial moment in the form:

$$\langle n(n-1)(t) \rangle \approx \frac{1}{\mu^2}t^2 + \frac{2}{\mu}\left(\frac{\alpha}{\mu^2} - 1\right)t + \frac{3\alpha^2}{2\mu^4} - \frac{\alpha}{\mu^2} + \frac{5}{6} \qquad (7)$$

for large $t$. Equations (6) and (7) then give the variance of the product number for large $t$:

$$\sigma_n^2(t) = \langle n(n-1)(t) \rangle + \langle n(t) \rangle - \langle n(t) \rangle^2 \approx \frac{\alpha}{\mu^3}t + \frac{1}{4}\left(\frac{5\alpha^2}{\mu^4} + \frac{1}{3}\right). \qquad (8)$$

From Eqs. (6) and (8), we can easily derive the Fano factor, i.e., the ratio of the variance to the mean of the product number in the steady state:

$$\lim_{t\to\infty} \frac{\sigma_n^2(t)}{\langle n(t)\rangle} = \frac{\alpha}{\mu^2}. \tag{9}$$

The Fano factor of the product number is related to the probability density function of the time required to produce at least $n$ products. As the product number $n$ approaches infinity, the limiting value of $n\frac{\sigma_{t_n}^2}{\langle t_n\rangle^2}$, proportional to the product number $n$ multiplied by the squared coefficient of the variation of the time to produce $n$ products, is expressed as

$$\lim_{n\to\infty} n\frac{\sigma_{t_n}^2}{\langle t_n\rangle^2} = \frac{\alpha}{\mu^2}, \tag{10}$$

where Eq. (1) has been used. Comparison of Eqs. (9) and (10) leads to the following equality:

$$\lim_{t\to\infty} \frac{\sigma_n^2(t)}{\langle n(t)\rangle} = \lim_{n\to\infty} n\frac{\sigma_{t_n}^2}{\langle t_n\rangle^2}. \tag{11}$$

If an event follows a renewal process, Eq. (11) reduces to the well-known existing result [12, 13]. Equation (11) is thus the generalization of the relation between event counting statistics and waiting time distribution in the steady state.

### 3. Examples of non-renewal processes

The non-renewal quotient [14], defined as $\left(\hat{\varphi}_n(s)/\hat{\varphi}_1^n(s)\right)-1$, determines how a stochastic process deviates from a renewal process. $\hat{\varphi}_n(s)$ represents the Laplace transform of $\varphi_n(t)$, which is the probability density function of the time required to complete $n$ events. A renewal process is characterized by $\hat{\varphi}_n(s)/\hat{\varphi}_1^n(s) = 1$ because all the time intervals between events in a renewal process have the same probability density function. On the contrary, the difference between $\hat{\varphi}_n(s)$ and $\hat{\varphi}_1^n(s)$, i.e., $\hat{\varphi}_n(s)/\hat{\varphi}_1^n(s) \neq 1$, signifies a non-renewal process. Here, three examples of non-renewal processes are considered, and the relevance of Eq. (11) for these processes is verified.

We first consider a non-renewal process whose inter-event time distribution is periodic. For example, suppose that a stochastic process repeats $q+1$ events, and an event occurs $q(\geq 1)$ times with the inter-event time distribution $\varphi_1(t)$, followed by another event occurring once with inter-event time distribution $\varphi_2(t)$. In this case, the Laplace transform of the probability density function of the time to complete $n$ events is given by

$$\hat{\varphi}_n(s) = \hat{\varphi}_1(s)^{n-\left[\frac{n}{q}\right]} \hat{\varphi}_2(s)^{\left[\frac{n}{q}\right]}, \tag{12}$$

where $[x]$ denotes the greatest integer less than or equal to $x$. Substitution of Eq. (12) into Eqs. (4-1) and (4-2) results in the left-hand side of Eq. (11) while the first and the second derivatives of Eq. (12) with respect to $s$, i.e., $\langle t_n^m \rangle = (-1)^m \left.\frac{\partial^m \hat{\varphi}_n(s)}{\partial s^m}\right|_{s=0}$, give the right-hand side of Eq. (11). Considering that $\langle t_k \rangle$ and $\langle t_k^2 \rangle$ denote the mean and the second moment of $\varphi_k(t)$ for $k = 1$ and 2, we obtain the steady-state statistics for Eq. (12) and verify that Eq. (11) is correct:

$$\lim_{t \to \infty} \frac{\sigma_n^2(t)}{\langle n(t) \rangle} = \lim_{n \to \infty} n \frac{\sigma_{t_n}^2}{\langle t_n \rangle^2} = \frac{(1+q)\left(\sigma_2^2 - \sigma_1^2 q\right)}{\left(\langle t_2 \rangle + \langle t_1 \rangle q\right)^2} \tag{13}$$

where $\sigma_1^2$ and $\sigma_2^2$ are the variances of the random variables having the probability density functions $\varphi_1(t)$ and $\varphi_2(t)$, respectively. It is clear from Eq. (13) that when $q$ vanishes, the periodic non-renewal process becomes a renewal process with the inter-event distribution $\varphi_2(t)$.

As the second example of a non-renewal process, we consider the process where the mean time between events increases with the number of times the event occurs. Suppose that the event occurs $n$ times during the time interval between the $(n-1)$-th event and the $n$th event. The number of events required for completing the $n$th event is given by $\sum_{m=1}^{n} m = n(n+1)/2$. Then, the Laplace transform of the probability density function of the time to complete $n$ events reads

$$\hat{\varphi}_n(s) = \hat{\varphi}_1(s)^{n(n+1)/2}, \tag{14}$$

whose derivatives with respect to $s$, together with Eqs. (4-1) and (4-2), lead us to prove the validity of Eq. (11) by the following expression:

$$\lim_{t\to\infty}\frac{\sigma_n^2(t)}{\langle n(t)\rangle} = \lim_{n\to\infty}\frac{n\sigma_{t_n}^2}{\langle t_n\rangle^2} = 0. \tag{15}$$

The third illustration of a non-renewal process is that the mean time between events decreases with the number of events that have occurred. The probability density function of the waiting time until the $n$th event occurs is modeled:

$$\hat{\varphi}_n(s) = \hat{\varphi}_1(s)^{\frac{2n}{1+n}}. \tag{16}$$

The comparison of the counting statistics and the waiting time statistics proves that Eq. (11) is correct:

$$\lim_{t\to\infty}\frac{\sigma_n^2(t)}{\langle n(t)\rangle} = \lim_{n\to\infty}\frac{n\sigma_{t_n}^2}{\langle t_n\rangle^2} = \infty. \tag{17}$$

A detailed consideration of Eqs. (14) and (16) reveals that the probability density function of the time to complete $n$ events is of the form $\hat{\varphi}_n(s) = \hat{\varphi}_1(s)^{nf(n)}$, where $f(n)$ is a function of $n$. As far as probability density functions of this form are concerned, a useful result is obtained by generalizing the right-hand side of Eq. (11):

$$\lim_{n\to\infty}\frac{n\sigma_{t_n}^2}{\langle t_n\rangle^2} = \frac{\sigma_{t_1}^2}{\langle t_1\rangle^2}\lim_{n\to\infty}\frac{1}{f(n)}, \tag{18}$$

where $\langle t_1\rangle$ and $\sigma_{t_1}^2$ are the mean and the variance, respectively, of the time distribution $\varphi_1(t)$. The validity of Eq. (18) is also confirmed by Eqs. (15) and (17). Note also that for $f(n) = 1$, Eq. (18) reduces to the result of the renewal theory.

## 4. Summary and discussion

We have presented in Eq. (11) the general relationship between the event counting statistics and the inter-event time statistics in the steady state, derived from Eqs. (4-1) and (4-2). Specifically, it is shown that the Fano factor, given by the variance divided by the mean, of the number of events occurring in the steady state equals the number of events multiplied by the squared coefficient of variation of the time to complete $n$ events in the limit of large $n$. This general relationship reduces to the well-known result for renewal processes. In addition, three examples of non-renewal processes have been examined to prove the validity of Eq. (11).

A variety of events in natural and social phenomena undergo a non-renewal process. It has been reported that some time series data cannot be explained by the renewal theory [1-11]. However, the analytical formula responsible for the stochastic properties of these data is unknown. In this paper, we have revealed the relationship between the event counting statistics and the inter-event time statistics for non-renewal processes in the steady state. In particular, Eq. (11) implies that we can extract one of the two statistics, the event counting and the interevent time statistics, without measuring both. The generalization of the relationship between the two, as described by Eq. (11), is a significant result and would broaden our understanding of non-renewal processes.

The methodology of this work can also be extended to a birth-death process. Since both birth and death are renewal processes, the relationship between the two statistics of the product number and creation/annihilation time is known [3, 15, 16]. However, in several real-world situations, the birth and death processes are non-renewal processes in which the probability density functions of the birth and death times vary with the occurrence of each birth and death [17-21]. The derivation of Eq. (11) is applicable to the derivation of the relationship between the product number statistics and the birth and death time statistics in the steady state of the non-renewal birth-death process. The study of this process is left for future work.

## Acknowledgment

This work was supported by the National Research Foundation (NRF) grants funded by the Korean government (Ministry of Science and ICT) [Grant No. 2021R1C1C2010450 (SJP) and No. 2022R1A2C1012532 (MYC)].

## Author Declarations

### Conflicts of interest

The authors have no conflicts of interest to declare.

### Data Availability

No data was used for the research described in this article.